\documentclass[twocolumn,aps,floatfix]{revtex4}

\usepackage{graphicx,epsfig,subfig,dcolumn,bm,mathrsfs,amsmath,amssymb}
\usepackage{color}
\begin{document}

\title{Stability of Two-Dimensional Soft Quasicrystals}

\author{Kai Jiang}
\email{kaijiang@xtu.edu.cn}
\affiliation{
School of Mathematics and Computational Science, Xiangtan University,
 Xiangtan 411105, P.R.~China}
\affiliation{ LMAM, CAPT and School of Mathematical Sciences, Peking University, Beijing 100871, P.R.~China}

\author{Jiajun Tong}
\affiliation{
 LMAM, CAPT and School of Mathematical Sciences, Peking University, Beijing 100871, P.R.~China}

\author{Pingwen Zhang}
\email{pzhang@pku.edu.cn}
\affiliation{
 LMAM, CAPT and School of Mathematical Sciences,
 Peking University, Beijing 100871, P.R.~China}

\author{An-Chang Shi}
\email{shi@mcmaster.ca}
\affiliation{
 Department of Physics \& Astronomy, McMaster
 University, Hamilton, Ontario Canada L8S 4M1}


\begin{abstract}
The relative stability of two-dimensional soft quasicrystals is examined using a recently developed projection method which provides a unified numerical framework to compute the free energy of periodic crystal and quasicrystals. Accurate free energies of numerous ordered phases, including dodecagonal, decagonal and octagonal quasicrystals, are obtained for a simple model, {\it i.e.}~the Lifshitz-Petrich free energy functional, of soft quasicrystals with two length-scales. The availability of the free energy allows us to construct phase diagrams of the system, demonstrating that, for the Lifshitz-Petrich model, the dodecagonal and decagonal quasicrystals can become stable phases, whereas the octagonal quasicrystal stays as a metastable phase.  
\end{abstract}


\maketitle

\section{Introduction}
\label{sec:intrd}

Quasicrystals are a class of ordered materials possessing quasiperiodic positional order {\em and} long-range orientational order. The intriguing property of quasicrystals is that their symmetry does not belong to the traditional crystallographic point groups. As such, the discovery of quasicrystals has led to a redefinition of the term crystal as any ordered materials having an essentially discrete diffraction pattern. For a traditional crystal or periodic crystal, the discrete diffraction pots form an periodic array on their reciprocal lattice. For a quasicrystal, the discrete diffraction spots densely fill the reciprocal space although, in practice, only the most intense reflections will be observed. Since the discovery of quasicrystals by Shechtman\,\cite{shechtman1984metallic} in a rapidly-quenched Al-Mn alloy thirty years ago, materials with quasicrystalline order have been found in more than a hundred different metallic alloys\,\cite{tsai2008icosahedral, steurer2004twenty}. Besides the large number of quasicrystals found in metallic alloys, quasicrystalline order has been observed in many soft condensed matter systems, including micelle-forming liquid crystals\,\cite{zeng2004supramolecular,percec2009self,percec2009self2}, block copolymers\,\cite{hayashida2007polymeric, zhang2012dodecagonal}, colloidal suspensions\,\cite{fischer2011colloidal} and binary mixtures of nanoparticles\,\cite{talapin2009quasicrystalline}. In contrast to the metallic quasicrystals, the building blocks of soft-matter quasicrystals are on a much larger length-scale, e.g.~tens to hundreds of nanometers. Soft quasicrystals provide an interesting platform for the study of fundamentals of quasiperiodic long-range orders, as well as potential candidates for advanced applications based on their unique electronic or photonic properties.

Due to a large number of studies since the discovery of quasicrystals, their structure and symmetry are now well understood\,\cite{janot1992quasicrystals, steurer2009crystallography}. In particular, quasicrystalline order can be described as the projection from a higher-dimensional periodic lattice\,\cite{hiller1985crystallographic}. On the other hand, the study of the thermodynamic stability of quasicrystals requires the examination of their free energy, which remains a challenge\,\cite{henley2006discussion, lifshitz2007soft}.

Theoretical approaches to investigating the origin and stability of an ordered phase, including periodic crystals and quasicrystals, often involve minimizing an appropriate free energy functional of the system, and comparing the free energies of different candidate structures\,\cite{chaikin1995principles,archer2013quasicrystalline}.  Therefore a systematic examination of the stability of quasicrystals requires the availability of suitable free energy functionals and accurate methods to compute the free energy of phases with quasicrystalline order. Several microscopic models have been developed over the years to explore ordered structures, which have yielded surprisingly rich phase diagrams. In some cases even stable quasicrystals are found\,\cite{denton1998stability, dzugutov1993formation,quandt1999formation, skibinsky1999quasicrystals, engel2007self,engel2008structural, keys2007quasicrystals, barkan2014controlled}. Parallel to the development of microscopic models, phenomenological theories based on coarse-grained free energy functionals have been widely used to study phases and phase transitions of ordered systems.  The utilization of such Landau-type theories provides an effective method to investigate the phase behaviour of physical systems exhibiting ordered phases. In the case of block copolymers, a free energy functional developed by Leibler\,\cite{leibler1980} has provided useful insight and a rather accurate description of the ordered phases of block copolymers. Similarly, for soft quasicrystals, a number of coarse-grained model free energy functionals have been proposed to explore the quasicrystalline order arising from model systems with more than one characteristic length-scales\,\cite{lifshitz1997theoretical, barkan2011stability,archer2013quasicrystalline, dotera2007mean,mermin1985mean, silber2000two, barkan2014controlled}. Among these models, the free energy functional proposed by Lifshitz and Petrich, or the Lifshitz-Petrich (LP) model\,\cite{lifshitz1997theoretical}, may be the simplest model with two length-scales. Initially the LP model was developed to study parametrically-excited surface waves (Faraday wave\,\cite{edwards1993parametrically}). Using a two-mode approximation, Lifshitz and Petrich obtained stable two-dimensional patterns with two-, six- and twelve-fold symmetries for the LP model. Recently the LP model has been used to explain the source of stability of soft quasicrystals with dodecagonal symmetry. In particular, the stability of soft quasicrystals is attributed to the existence of two length-scales and three-body interactions in the free energy functional\,\cite{lifshitz2007soft}. Nevertheless, the existence and stability of other quasiperiodic structures within the LP model remain an open question.

Besides a proper free energy functional of the system, the study of thermodynamic stability requires methods to compute the free energy of the different ordered phases accurately. Due to the spatial periodicity, the computation of periodic crystals can be carried out within a unit cell with periodic boundary conditions. On the contrary, it is not possible to reduce the structure of a quasicrystal to unit cells because quasicrystals are space-filling ordered structure without spatial periodicity. In the literature, a common method to overcome this difficulty is to utilize periodic structures with large unit cells to approximate quasicrystals\,\cite{quandt1999formation, skibinsky1999quasicrystals, engel2007self, lifshitz1997theoretical,reinhardt2013computing, dotera2006dodecagonal}; their free energies are also approximated by that of these periodic orders as well.  However, it is not immediately obvious whether such approximation of the free energy is accurate. Indeed, Jiang and Zhang\,\cite{jiang2014numerical} recently showed that the free energy density of dodecagonal quasicrystal is {\em always} lower than its approximant evaluated by this approximated approach.

An alternative approach to calculate the free energy of quasicrystals is based on the observation that quasiperiodic lattices can be generated by a cut-and-project method from higher-dimensional periodic lattices\,\cite{janot1992quasicrystals}. It follows that the density and free energy of quasicrystals can be obtained using the quasiperiodic lattices derived from the higher-dimensional periodic structure. A previous approach along this line is the Gaussian method, in which the density profile of a quasicrystal is assumed to be given by a sum of Gaussian functions centered at the lattice points of a predetermined quasicrystalline lattice\,\cite{mcarley1994hard}. The width of the Gaussian function is treated as a variational parameter, which is optimized to minimize the free energy of the system. More recently, Jiang and Zhang\,\cite{jiang2014numerical} have proposed a generalized spectral method for characterizing the density profile and computing the free energy of quasiperiodic structures.  Instead of working in the real space, Jiang and Zhang utilized the fact that the Fourier spectrum of quasicrystalline structures can be lifted into that of a higher-dimensional periodic structure, so that the Fourier spectrum of the quasiperiodic pattern can be obtained by projecting the higher-dimensional reciprocal lattice vectors onto the original Fourier space through a projection matrix.  Details of the lifting and projection depend on the symmetry and rank of the desired structures. As a special case, the projection method can also be used to investigate periodic crystals by setting the projection matrix as an identity matrix. From this perspective the projection method provides a unified computational framework for the study of periodic crystals and quasicrystals. In particular, the free energy of periodic crystals and quasicrystals can be obtained with the same accuracy.

In the current work, we apply the projection method to study the relative stability of different ordered phases of the LP model. The main objective is to investigate the existence and relative stability of periodic structures and quasicrystals beyond the 12-fold symmetric patterns. Specifically, two-dimensional quasiperiodic patterns with 12-, 10- and 8-fold symmetries as well as several periodic crystal structures including the three-dimensional body-centered-cubic (BCC) phase (with space group $Im\bar{3}m$) have been obtained as possible phases of the LP model. A comparison of the free energies of these candidate phases leads to the construction of a phase diagram of the system, thus extending the earlier results of Lifshitz and Petrich\,\cite{lifshitz1997theoretical}. In particular, the theoretical study predicts that the decagonal quasicrystal can become a stable phase. 

\section{Theoretical Framework}
\label{theory}

Since the discovery of quasicrystals, a large number of theoretical studies have been carried out to investigate their symmetry, structure characterization and thermodynamic properties. In general, the formation of quasicrystalline phases is characterized by two or more length-scales. One possibility of realizing two length-scales is to have a pairwise interaction that depends on two length-scales explicitly. For example, a recent Monte Carlo simulation study by Dotera {\it et al.}\,\cite{dotera2014} demonstrated that a simple step-like interaction potential with two length-scales could lead to the formation of various two-dimensional quasicrystals. Another possibility of realizing two length-scales is to construct a Landau-type free energy functional with two or more characteristic length-scales\,\cite{lifshitz2007soft, barkan2011stability, archer2013quasicrystalline}. In particular, Lifshitz and Petrich\,\cite{lifshitz1997theoretical} have proposed a Landau theory with a scale order parameter and with a second-order term containing two length-scales explicitly. It has been demonstrated that the LP model exhibits dodecagonal quasicrystal. In what follows we will use the LP theory as a model free energy functional to examine the relative stability of various two-dimensional quasicrystals.

In order to examine the relative stability of different ordered phases, the free energies of the candidate structures have to be evaluated accurately. The candidate phases correspond to free energy local minima in the free energy landscape of the system.  Therefore the computation of their free energies corresponds to finding the solutions of the Euler-Lagrange equation of the free energy functional. For quasiperiodic phases, a direct computation in the real space can only be carried out approximately using periodic approxmants or other truncation methods. On the other hand, the computation can be carried out in reciprocal or Fourier space using projection method as described recently by Jiang and Zhang\,\cite{jiang2014numerical}. With appropriate choice of the Fourier basis, the projection method provides a unified numerical framework to study periodic crystals and quasicrystals with the same accuracy.  

In this work we apply the projection method to the LP model and determine the relative stability of different phases including two-dimensional dodecagonal, decagonal and octagonal quasicrystals. The rest of this section gives a brief introduction to the LP model and the projection method for ordered phases.

\subsection{Lifshitz-Petrich model}
\label{sec:lpmodel}
The Lifshitz-Petrich model is a generic Landau theory for a physical system with two length-scales\,\cite{lifshitz1997theoretical}. The original LP model was motivated by experiments on the parametrically-excited surface waves, which exhibits dodecagonal quasiperiodic order\,\cite{edwards1993parametrically}. Lifshitz and Petrich\,\cite{lifshitz1997theoretical} introduced a spatial-varying scale order parameter, $\phi(\mathbf{r})$, and constructed a free energy functional, $F_{\mathrm{LP}}[\phi(\mathbf{r})]$, which governs the pattern-forming dynamics of the system. The essential characteristic of the LP model is that the disordered phase, corresponding to $\phi(\mathbf{r})=0$, undergoes an instability simultaneously at two wave-lengths. Specifically the Lifshitz-Petrich model is defined by the free energy functional,
\begin{align}
	\nonumber
	F_{\mathrm{LP}}[\phi(\mathbf{r})] = \,\frac{1}{V}\int
	d\mathbf{r}\,\Big\{\frac{c}{2}&[(\nabla^2+1)(\nabla^2+q^2)\phi]^2
	\\
	&-\frac{\varepsilon}{2}\phi^2-\frac{\alpha}{3}\phi^3+\frac{1}{4}\phi^4\Big\},
	\label{eqn:LP}
\end{align}
where $V$ is the volume of the system, $c$ is an energy penalty factor, $\varepsilon$ is a temperature-like controlling parameter, and $\alpha$ is a parameter characterizing the third-order interactions. The order parameter $\phi(\mathbf{r})$ corresponds to the density profile for a soft matter system.  Comparing with the usual Landau theory of phase transition, the crucial feature of the LP model is the occurrence of the second length-scale specified by the parameter $q$. It should be noticed that the basic length scale of the model is take as the unit wavelength (as specified by the $\nabla^2+1$ term) and the second length-scale is specified by a wavevector, $q$, which is the ratio of the two characteristic length-scales. In the case of the surface waves, the origin of the two length-scales is the two-frequency parametric excitation of the fluid surface\,\cite{lifshitz1997theoretical}. In the case of generic soft matter system, the two length-scales could be from the pairwise interactions of the soft particles.  In the current work the parameter $q$ is taken as one of the controlling parameters of the systems.  In principle, free energy minimization should be carried out with respect to $q$ as well. However, carrying out such minimization is a formidable task. Instead, we study the relative stability of different patterns with a limited number of $q$-values.

In their original study\,\cite{lifshitz1997theoretical}, Lifshitz and Petrich used a two-mode approximation to describe the different ordered phases. They demonstrated that the simple LP model exhibits various two-dimensional ordered patterns with 2-, 4-, 6- and 12-fold symmetries, and obtained a phase diagram for the LP model based on the two-mode approximation. Subsequent studies\,\cite{lifshitz2007soft} have demonstrated that such free energy functionals with two length-scales could be considered as a generic model for self-assembling soft matter systems, providing insights into the essential features of soft quasicrystals forming in dendrimers, tetrablock terpolymers and micelle-forming liquid crystals. Furthermore, the theory with two length-scales has been used to study the soft-particle systems\,\cite{barkan2011stability, archer2013quasicrystalline, barkan2014controlled}.  It is also noticed that three-body interactions characterized by the cubic term in the LP free energy functional plays an important role in stabilizing quasicrystals.

\subsection{Projection Method}
\label{sec:pm}
For a given set of parameters, the candidate phases are local minima of the free energy functional, corresponding to solutions of the Euler-Lagrange equation,
\begin{align}
\frac{\delta F_{\mathrm{LP}}}{\delta\phi(\mathbf{r})}=0.
\label{eqn:el}
\end{align}
For the LP model, the Euler-Lagrange equation (Eq.~\ref{eqn:el}) corresponds to a eighth-order nonlinear partial differential equation, which can be solved using a number of numerical methods. For structures with quasicrystalline order, an efficient method is the projection method formulated in the Fourier space.  It is based on the observation that the Fourier spectrum of a $d$-dimensional aperiodic structure, such as a quasicrystal, consists of Bragg peaks on a, $n$-dimensional lattice, where $n\geqslant d$. In other words, the Fourier spectrum of the quasiperiodic structure can be lifted into an $n$-dimensional periodic lattice. The $n$-dimensional reciprocal vectors can be spanned by a set of bases $\mathbf{b}_i$, which are the primitive reciprocal vectors in the $n$-dimensional reciprocal space, with integer coefficients, {\em i.e.}~$n$-dimensional reciprocal vector $\mathbf{H}$ can be written as $\mathbf{H}=\sum_{i=1}^n h_i \mathbf{b}_i\in\mathbb{R}^n$, $h_i\in\mathbb{Z}$,
$\mathbf{b}_i \in\mathbb{R}^n$ \cite{janssen2007aperiodic,steurer2009crystallography}.  The physical $d$-dimensional wavevector $\mathbf{k}$ is then obtained from the $n$-dimensional vector $\mathbf{H}$ by a projection, $\mathbf{k}=\mathcal{S}\cdot\mathbf{H}$, where $\mathcal{S}$ is a projection matrix of $d\times n$-order. The dimensionality $n$ and the specific form of the projection matrix is determined by the structure of the ordered phases\,\cite{steurer2009crystallography}. The expression of $\mathcal{S}$ is not unique and relies on the symmetry of the quasicrystals. More precisely, it depends on the choice of the basis vectors.

Using the $n$-dimensional vectors and the projection operator, the Fourier expansion of any quasiperiodic function $\phi(\mathbf{r})$ can be written in the form,
\begin{align}
	\phi(\mathbf{r}) =
	\sum_{ \{\mathbf{k}\} }\hat{\phi}_\mathbf{k}e^{i \mathbf{k} \cdot \mathbf{r}}.
	\label{eqn:pm}
\end{align}
In the above expression, $\mathbf{r}\in\mathbb{R}^d$, and $\mathbf{k}=\sum_{i=1}^n h_i (\mathcal{S} \mathbf{b}_i)  \in\mathbb{R}^d$. One simple observation is that the expansion (Eq.~\ref{eqn:pm}) allows us to treat the $d$-dimensional quasiperiodic structure as a slice of an $n$-dimensional periodic structure whose orientation is determined by $\mathcal{S}$.

Expanding the order parameter $\phi(\mathbf{r})$ in the form of Eq.~\ref{eqn:pm} and inserting it into Eq.~\ref{eqn:LP}, the free energy functional can be written in terms of the Fourier coefficients $\hat{\phi}_\mathbf{k}$. For a given structure of interest, the reciprocal lattice vectors are determined by its symmetry, and the optimal coefficients are obtained by minimizing the free energy functional.  In our previous work\,\cite{jiang2014numerical}, it has been shown that, when using the projection method, it suffices to have a free energy functional defined in the lower (physical) $d$-dimensional space.  Therefore the computations are implemented in the $n$-dimensional space, while the final results represent the $d$-dimensional structures through (\ref{eqn:pm}). The computation in the $n$-dimensional space is carried out on a regular periodic grid since the reciprocal lattice in this space is periodic. As a special case of quasiperiodic structures, a $d$-dimensional periodic structure can be described within the projection method by setting the projection matrix as a $d\times d$ identity matrix. In this case the projection method is reduced to the commonly used Fourier-spectral method. This dose not provide any computational advantage when it comes to periodic crystals.  However, this view provides a unified computational scheme of the periodic crystals and quasicrystals.

In practice, we adopt a dissipative method to obtain solutions of the Euler-Lagrange equation of the free energy functional.  Inserting the generalized Fourier expansion (Eq.~\ref{eqn:pm}) into the LP model, the free energy functional (Eq.~\ref{eqn:LP}) becomes a function of the Fourier coefficients,
\begin{equation}
	\begin{aligned}
		F_{\mathrm{LP}} &=
		\frac{1}{2}\sum_{\mathbf{k}_1+\mathbf{k}_2=0}
		\left\{c \Big(1-\sum_{k=1}^d g_k^2\Big)^2
		\Big(q^2-\sum_{k=1}^d g_k^2\Big)^2-\varepsilon\right\}
	\\
	&\times \hat{\phi}(\mathbf{k}_1)\,\hat{\phi}(\mathbf{k}_2)
	 -\frac{\alpha}{3}\sum_{\mathbf{k}_1+\mathbf{k}_2+\mathbf{k}_3=0}\hat{\phi}(\mathbf{k}_1)\,\hat{\phi}(\mathbf{k}_2)\,\hat{\phi}(\mathbf{k}_3)
	\\
	&+\frac{1}{4}\sum_{\mathbf{k}_1+\mathbf{k}_2+\mathbf{k}_3+\mathbf{k}_4=0}
	\hat{\phi}(\mathbf{k}_1)\,\hat{\phi}(\mathbf{k}_2)\,\hat{\phi}(\mathbf{k}_3)\,\hat{\phi}(\mathbf{k}_4),
	\end{aligned}
	\label{eqn:LPqcFour}
\end{equation}
where $g_k = \sum_{i=1}^n \sum_{j=1}^n s_{ki}h_j b_{ji}$, $k=1,2,\dots,d$, $s_{ki}$ are the components of the projection matrix $\mathcal{S}$, and $b_{ji}$ is the $i$-th component of the primitive reciprocal vector $\mathbf{b}_j$. Instead of solving the nonlinear Euler-Lagrange equation directly, we adopt a relaxation method to solve the minimization problem.  Specifically, the Fourier coefficients are iterated according to the following relaxation equation,
\begin{align}
	\nonumber
	\frac{\partial \hat{\phi}_\mathbf{k}}{\partial t} =
	& \,
	\varepsilon \hat{\phi}_\mathbf{k}
	+ \alpha \hat{\phi}_\mathbf{k}^2 - \hat{\phi}_\mathbf{k}^3
	\\
	&-c \Big(1-\sum_{k=1}^d g_k^2\Big)^2
		 \Big(q^2-\sum_{k=1}^d g_k^2\Big)^2
	\hat{\phi}_\mathbf{k}.
	\label{eqn:LP:PM}
\end{align}
It should be pointed out that the variable $t$ here is not time, but a parameter controlling the iteration steps. In this expression the quadratic term and cubic term are given by,
\begin{equation*}
	\begin{aligned}
	\hat\phi^2_{\mathbf{k}}&=\sum_{\mathbf{k_1}+\mathbf{k}_2=\mathbf{k}}
\hat{\phi}_{\mathbf{k}_1}\hat{\phi}_{\mathbf{k}_2},
	\\
	\hat\phi^3_{\mathbf{k}}&=\sum_{\mathbf{k}_1+\mathbf{k}_2+\mathbf{k}_3=
\mathbf{k}} \hat{\phi}_{\mathbf{k}_1} \hat{\phi}_{\mathbf{k}_2}
\hat{\phi}_{\mathbf{k}_3}.
	\end{aligned}
	\label{eqn:LP:nonlinear}
\end{equation*}
From these expressions it is obvious that the nonlinear (quadratic and cubic) terms in Eq.~\ref{eqn:LP:PM} are $n$-dimensional convolutions in the reciprocal space. A direct evaluation of these nonlinear terms will be computationally expensive.  Instead, these terms are simple multiplication in the $n$-dimensional positional space and the computation of these nonlinear terms in the position space is straightforward. The pseudospectral method takes advantage of this observation by evaluating the gradient terms in the Fourier space and the nonlinear terms in the position space, thus providing an efficient technique to find solutions of the Euler-Lagrange equation. The pseudospectral method requires access to the density function in real and reciprocal spaces. The transformation between the real-space and reciprocal space was done by performing Fast Fourier Transformation (FFT) in the $n$-dimensional space.

Starting from an initial configuration with a specified symmetry, a steady state solution of Eq.~\ref{eqn:LP:PM}, corresponding to a local minimum of the free energy functional, is obtained. Using initial configurations with different symmetries leads to different ordered structures as solutions of the minimization problem. The ordered structures corresponding to these solutions are taken as candidate phases of the problem. The free energies of these candidate structures are then compared and used to construct phase diagrams of the system.

\section{Results and Discussion}
\label{sec:rslt}

Using the projection method outline above, we will obtain possible ordered phases, and examine their relative stability, of the LP model. Due to the computational demand, in the current study, the quasicrystals of interest are restricted to two-dimensional quasiperiodic patterns whose point group symmetries can be realized by periodic lattices in 4-dimensional space, namely 5-, 8-, 10- and 12-fold symmetric structures\,\cite{hiller1985crystallographic}. Therefore the dimensions of the physical space and the reciprocal space are $d = 2$ and $n = 4$. Besides, one-, two- and three-dimensional periodic patterns, corresponding to the commonly observed lamellar, cylindrical and spherical phases, are included in our study. In practice, the $n$-dimensional Fourier space is discretized using $24$ basis functions along each direction. The total number of variables is thus $24^n$. We remark that the projection method works equally well if we use different numbers of basis functions along each dimension, although we will not do that here for simplicity.

A semi-implicit scheme is adopted to solve Eq.~\ref{eqn:LP:PM} until the relative change of $F_{\mathrm{LP}}$ between consecutive iteration steps is smaller than $10^{-8}$.  For the cases of periodic crystals and quasicrystals, the computation starts with initial configurations with the desired symmetries.  The choice of initial configurations will speed up the computations by reducing the number of iterations. If the initial symmetric structure is a local minimum of the free energy functional for a given set of model parameters, the calculation will lead to a converged solution with the prescribed symmetry. In the case that the chosen symmetry of the initial configuration does not correspond to a local minimum, the iteration procedure will lead to other phases, or more commonly, to the trivial solution $\phi(\mathbf{r})=0$ corresponding to the disordered phase. It should be noticed that the homogeneous phase is always a solution of the Euler-Lagrange equation. 

\subsection{Quasicrystals and Periodic Crystals from the LP Model}

In this subsection we enumarate the ordered phases used in our study. All these ordered phases are solutions of the Euler-Lagrange equation, corresponding to local minima of the LP free enegy functional. These ordered phases are used as candidate structures for the construction of the phase diagram. At a given point of phase space only one of these phases could become stable whereas all the others are metastable.  

\begin{figure}[t]
	\centering
		\includegraphics[scale=0.53]{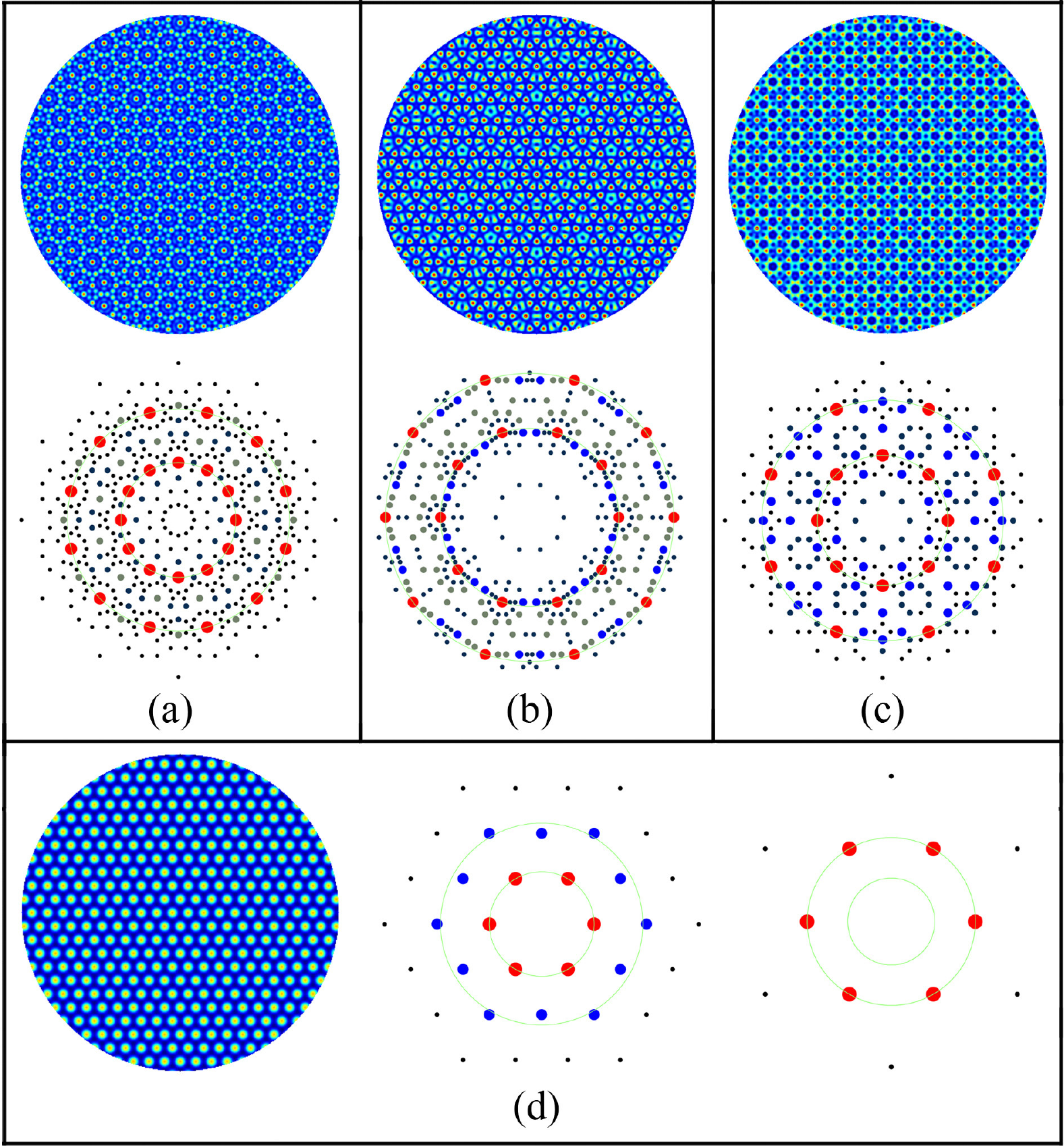}
	\caption{\label{fig:phases} Order parameter profiles
	$\phi(\mathbf{r})$ and $\hat\phi_{\mathbf{k}}$ (with
	$\hat\phi_{\mathbf{k}=0}$ removed) obtained
	by minimizing free energy density (Eq.~\ref{eqn:LP}) with different choices of $q$.
	For (a)-(c), the upper panels are density profiles, while the below ones are the corresponding Fourier transforms.
	(a) 12-fold symmetric pattern for $q=2\cos(\pi/12)$,
	(b) 10-fold symmetric pattern for $q=2\cos(\pi/5)$,
	(c) 8-fold symmetric pattern for $q=2\cos(\pi/8)$,
	(d) 6-fold symmetric crystal with two different length
	scales, located at $|\mathbf{k}|=1$ and $|\mathbf{k}|=q$ ($q\neq 1$).
	}
	\label{fig:phases}
\end{figure}
With proper choice of the second length-scale characterized by $q$, solutions corresponding to periodic and quasiperiodic patterns are obtained as local minima of the LP free energy functional. In particular, the choice of $q=2\cos(\pi/12)$, $2\cos(\pi/5)$ and $2\cos(\pi/8)$ leads to three two-dimensional quasicrystals, with 12-, 10- and 8-fold symmetries, as the potential stable phases in the model. The order parameter or density profiles $\phi(\mathbf{r})$, and the corresponding profiles in the Fourier space $\hat{\phi}_{\mathbf{k}}$, of these ordered structures are shown in Fig.\,\ref{fig:phases}(a)-(c). Besides these quasicrystalline phases, several periodic structures, corresponding to two-dimensional patterns with 2-, 4- and 6-fold symmetries and a three-dimensional periodic phase of spheres packed on a body-centered-cubic (BCC) lattice, have been observed as local minima of the LP free energy functional.  Fig.\,\ref{fig:phases}(d) presents the order parameter profiles of the 6-fold symmetric pattern. Because the energy penalty factor $c$ is finite, more nonzero Fourier modes which do not set on two rings of radius $1$ and $q$ appear as Fig.\,\ref{fig:phases} shows. Before we examine the stability of these ordered structures, it is useful to verify the symmetries of these quasicrystalline patterns.

\textit{Dodecagonal Quasicrystal (DDQC)} --- The 12-fold symmetric 
dodecagonal quasicrystal is obtained by choosing structural parameter $q=2\cos(\pi/12)$. The corresponding projection matrix is given by,
\begin{equation}
	\mathcal{S} =\left(
	\begin{array}{cccc}
		1 & \cos(\pi/6) & \cos(\pi/3) & 0 \\
		0 & \sin(\pi/6) & \sin(\pi/3) & 1
	\end{array}
\right).
\label{eqn:DDQC:projMatrix}
\end{equation}
As is shown in Fig.\,\ref{fig:phases} (a), the Fourier spectrum of this structure consists of two stars of reciprocal vectors as its principal Fourier components (red dots in Fig.\,\ref{fig:phases} (a)). Each star contains 12 vectors separated evenly by an angle of $30^\circ$. The reciprocal vectors on the $|\mathbf{k}| = q$ circle are sums of two neighboring reciprocal vectors on the $|\mathbf{k}| = 1$ circle. The real space and Fourier space profiles of this structure clearly exhibit 12-fold rotational symmetry. The structural parameter of the dodecagonal quasicrystal is in agreement with that of Lifshitz and Petrich\,\cite{lifshitz1997theoretical}, where the dodecagonal pattern was obtained with the same choice of $q$.

\textit{Decagonal Quasirystal (DQC)} --- The 10-fold symmetric
quasicrystal is obtained when $q$ is set at $q=2\cos(\pi/5)$, with a projection matrix specified by,
\begin{equation}
	\mathcal{S} =\left(
	\begin{array}{cccc}
		1 & \cos(\pi/5) & \cos(2\pi/5) & \cos(3\pi/5) \\
		0 & \sin(\pi/5) & \sin(2\pi/5) & \sin(3\pi/5)
	\end{array}
\right).
\label{eqn:DQC:projMatrix}
\end{equation}
Fig.\,\ref{fig:phases} (b) gives the order parameter profiles of this phase in the real space and Fourier space, both of which exhibit the 10-fold rotational symmetry. The principal Fourier components (red dots in Fig.\,\ref{fig:phases} (b)) contain twenty reciprocal vectors, with ten vectors located on the circle of $|\mathbf{k}| = 1$, and the other ten on the circle of $|\mathbf{k}| = q$. Unlike the dodecagonal quasicrystal, the reciprocal vectors of the decagonal quasicrystal located on the $|\mathbf{k}| = q$ and $|\mathbf{k}| = 1$ circles are collinear.

It should be noticed that the decagonal quasicrystal has the same diffraction pattern as the pentagonal quasicrystal\,\cite{senechal1996quasicrystals}. Therefore identification of the decagonal quasicrystal requires a detailed examination of the Fourier coefficients. For the LP free energy functional, it has been found that the Fourier coefficients on a given circle are real and equal to each other ({\it e.g.}~$\hat\phi_{\mathbf{k}=1}=0.759_2$, $\hat\phi_{\mathbf{k}=q}=0.694_6$ when $c=100$, $\alpha=10.0$ and $\varepsilon=0.5$). From this observation it can be concluded that the structure obtained from the calculations has a 10-fold symmetry, corresponding to decagonal quasicrystal.

\textit{Octagonal Quasicrystal (OQC)} --- The third aperiodic
structure is the 8-fold symmetric octagonal quasicrystal that is found by choosing $q=2\cos(\pi/8)$ and a projection matrix,
\begin{equation}
	\mathcal{S} =\left(
	\begin{array}{cccc}
		1 & \cos(\pi/4) & \cos(\pi/2) & \cos(3\pi/4) \\
		0 & \sin(\pi/4) & \sin(\pi/2) & \sin(3\pi/4)
	\end{array}
\right).
\label{eqn:OQC:projMatrix}
\end{equation}
As shown in Fig.\,\ref{fig:phases} (c), the principal Fourier coefficients of this structure compose of two 8-fold symmetric stars of wavevectors on the $|\mathbf{k}| = 1$ and $|\mathbf{k}| = q$ circles. Similar to the case of DDQC, the wavevectors located on $|\mathbf{k}| = q$ circle are sums of the neighboring wavevectors located on the $|\mathbf{k}| = 1$ circle.

\textit{Periodic Crystals} --- The projection method is used to obtain
periodic structures as local minima of the LP free energy functional by setting the projection matrix as a unit matrix. Due to the existence of two characteristic length-scales in the LP model, two stable periodic structures with the same symmetry but different lattice spacings can be obtained\,\cite{lifshitz1997theoretical,archer2013quasicrystalline, silber1998nonlinear}. These structures are termed \emph{sibling periodic crystals}. When $q\neq 1$, a number of periodic phases with their sibling periodic crystals, including two-dimensional 2-, 4- and 6-fold symmetric patterns and a three-dimensional BCC phase, have been obtained from our calculations. The basic Fourier vectors of two sibling phases are located at the circles with radii $|\mathbf{k}| =1$, and $|\mathbf{k}| = q$, respectively. For instance, Fig.\,\ref{fig:phases} (d) presents the real densities and the distinct Fourier spectra corresponding to two hexagonal sibling patterns. Note that we only show one copy of the real density, since those associated with sibling phases only differ in scale.

It should be noticed that the candidate structures obtained from our calculations differ from that obtained by Lifshitz and Petrich\,\cite{lifshitz1997theoretical}. Specifically, Lifshitz and Petrich reported that the DDQC could become an equilibrium phase with $q=2\cos(\pi/12)$, whereas the OQC or DQC were not obtained for any choice of $q$. This difference can be attributed to the different numerical methods used in the calculations. In Lifshitz and Petrich's simulations, a pseudospectral algorithm is applied to a large square computational box with periodic boundary condition. For a given quasicrystal, the periodic computational domain is strictly determined by the non-crystallographic rotational symmetry, yet still having some approximate error\,\cite{jiang2014numerical, rucklidge2009design}. On the other hand, the projection method employed in the current study respects the symmetry of the quasicrystals strictly, leading to a robust method to obtain quasiperiodic structures numerically.

\subsection{Relative Stability of Ordered Phases and Phase diagram}
\label{subsec:Relative Stability of Ordered Phases and Phase diagram}

For the LP free energy functional, three quasicrystals and four periodic crystals (and their siblings) are obtained as the candidate phases for the model system. The relative stability of these phases is obtained by comparing their free energies. A phase diagram in the $\varepsilon$-$\alpha$ plane can be constructed from the free energy comparison. In this section we will examine factors influencing the relative stability of the different phases first, and then present the phase diagram of the system.  The value of $c$ will affect the stable areas of different phases. Our extensive calculations for cases with different values of $c$ demonstrate that the boundaries between different phases do change with $c$. However the topological of the stable regions of the different phases is insensitive to the $c$-values. Therefore in what follows we will use $c=100$ as an example of finite $c$, and examine the phase behaviour of the system as a function of $\varepsilon$ and $\alpha$.  As a initial step in the study of the phase behaviour of the system, we will examine the relative stability of the sibling phases first. Results for the other ordered phases are presented subsequently.

\begin{figure}[b]
	\centering
		\includegraphics[scale=0.25]{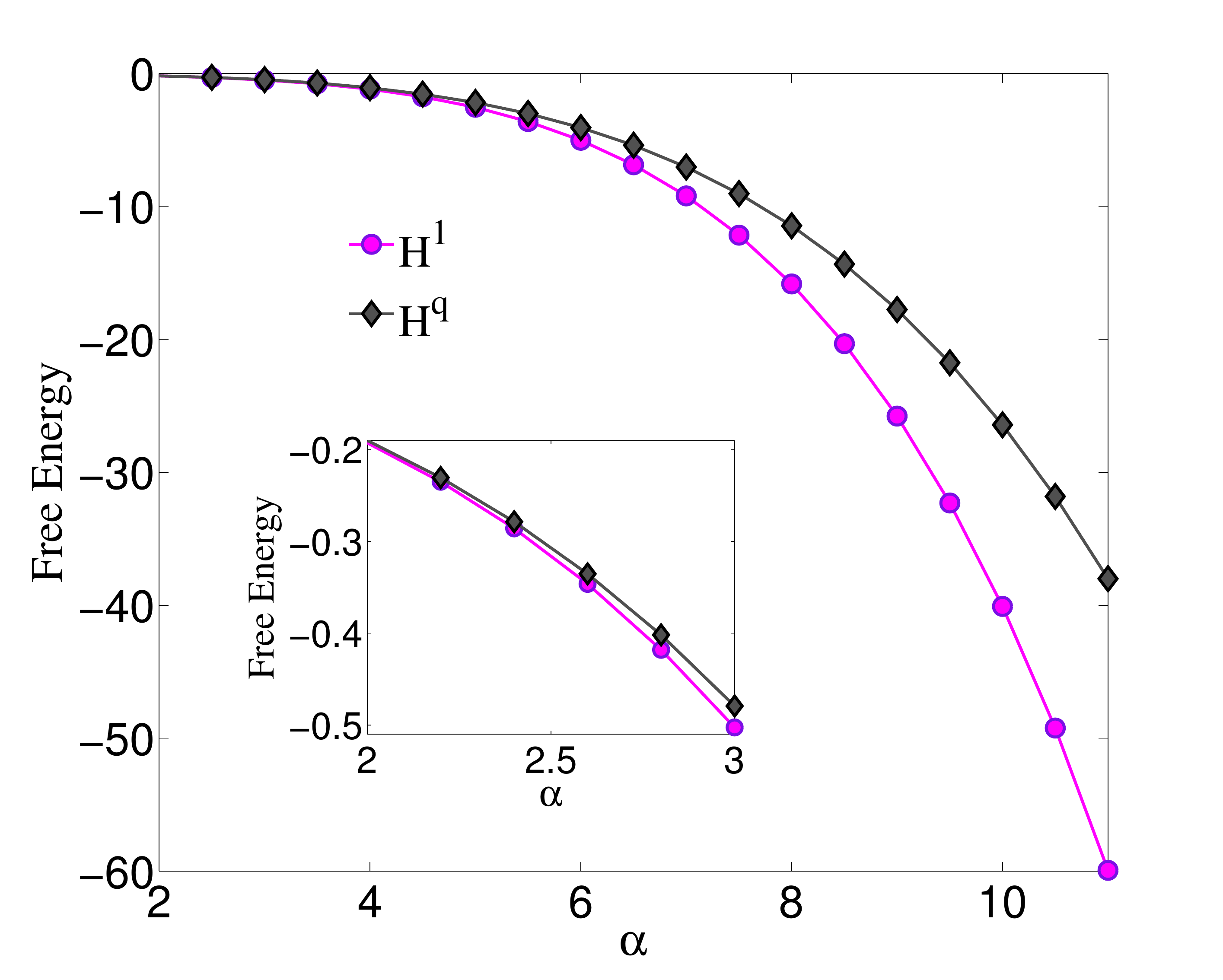}
		\caption{Free energy of
		the large hexagonal pattern $\mathrm{H}^1$ and
		the small hexagonal pattern $\mathrm{H^q}$ as a function of
	  $\alpha$ for fixed $\varepsilon=0.5$, $q=2\cos(\pi/5)$, and $c=100$.
	  }
	\label{fig:energy6fold10xi0.5}
\end{figure}

\textit{Relative Stability of Sibling Periodic Crystals} ---
Sibling periodic crystals are obtained by imposing the same symmetry, but with different length-scales, on the initial configurations of $\phi(\mathbf{r})$. As an example, Fig.\,\ref{fig:phases} (d) shows the real-space density profile and Fourier spectra of the hexagonal sibling phases. For convenience, the hexagonal structure with larger lattice spacing in the real space is termed as $\mathrm{H}^1$ (the middle figure in Fig.\,\ref{fig:phases} (d)), and the one with the smaller lattice spacing as $\mathrm{H}^q$ (the right figure in Fig.\,\ref{fig:phases} (d)).  Fig.\,\ref{fig:energy6fold10xi0.5} shows the free energy density (computed from Eq.~\ref{eqn:LP}) of the hexagonal structures for $\varepsilon=0.5$, $q=2\cos(\pi/5)$ as a function of $\alpha\in[2.0,11.0]$. It is observed that the free energy of $\mathrm{H}^1$ is always lower than that of $\mathrm{H}^q$. Our extensive calculations show that for other choices of $\varepsilon$, $c>0$ and $q\neq 1$, the free energy of $\mathrm{H}^1$ is always lower than that of $\mathrm{H}^q$. As $c$ increases, the free energy differences between the two hexagonal sibling phases decrease. When $c\rightarrow \infty$, the Fourier modes with nonzero coefficients are constrained to be on the circle with radius $1$ or $q$. In this case the free energies of the two hexagonal sibling crystals become equal. The comparison of the free energy of the two sibling HEX phases clearly demonstrate that the $\mathrm{H}^1$ structure always has lower free energy than the $\mathrm{H}^q$ phase over the range of model parameters used in our calculations.

For other sibling periodic crystals, such as the two-dimensional patterns with 2-fold symmetry and the three-dimensional BCC phases, we also find that the phases with larger lattice spacing in the physical space have lower free energy than the one with smaller lattice spacings. Therefore, only the periodic phases with larger lattice spacing may become equilibrium phases on the phase diagram. In the following we will only consider the phases with larger lattice spacing in the phase diagram. In particular, we will term the hexagonal phase $\mathrm{H}^1$ in the phase diagram as the HEX phase.  
\begin{figure}[b]
	\centering
		\includegraphics[scale=0.25]{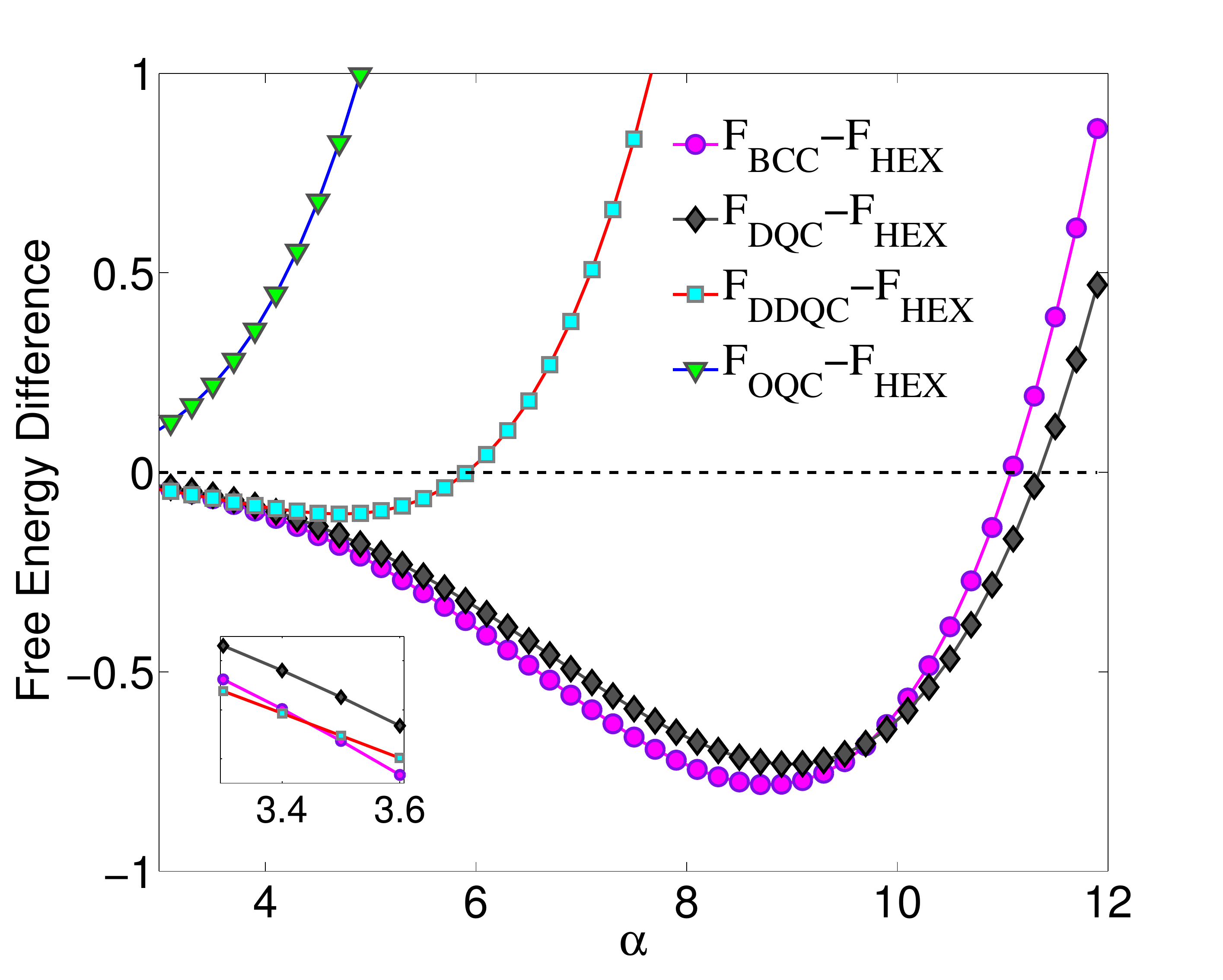}
	  \caption{The free energy differences from the value of
	  HEX of various structures as a function of $\alpha$
	  on the phase path of fixed $\varepsilon=-0.1$ and
	  $c=100$.
	  }
	\label{fig:energyplotxi-0.1}
\end{figure}

\textit{Relative Stability of Quasicrystals and Periodic Crystals} --- 
We now examine the relative stability of quasicrystals and periodic crystals by comparing their free energies in the $\varepsilon$-$\alpha$ plane. In what follows we will treat $q$ as one of controlling parameters of the system, dependent on the symmetry of ordered phases, that is chosen a priori.  For the quasiperiodic phases, the choice of $q$ for the three two-dimensional quasicrystals of interest have been given before.  For the periodic crystals, different values of $q$ may affect the free energy of the sibling periodic crystals. However, it will not significantly change the relative stability of quasicrystals and periodic crystals, merely affects the stability region.  Therefore, in what follows we set $q=2\cos(\pi/5)$ for periodic crystals.

From our calculations, it is found that the 4-fold symmetric pattern is always metastable.  Therefore, this structure is not included in the following analysis. The candidate structures are therefore composed of the three quasicrystals (DDQC, DQC and OQC), the HEX phase and the BCC phase. Using the free energy of the HEX phase as the baseline, the free energies of the DDQC, DQC, OQC and BCC phases as a function of $\alpha$ are shown in Fig.\,\ref{fig:energyplotxi-0.1} with $\varepsilon=-0.1$.  From the results shown in Fig.\,\ref{fig:energyplotxi-0.1}, it is obvious that the OQC has higher free energy than the other four candidate structures. Thus the OQC structure is a metastable phase of the system with the given set of parameters. On the other hand, the other four candidate structures (DDQC, DQC, HEX and BCC) can become equilibrium phases with the lowest free energy. Specifically, when the value of $\alpha$ is increased, the phase transition sequence is predicted to be DDQC to BCC to DQC to HEX. The corresponding stable regions of these phases are $\alpha\leq 3.44$ (DDQC), $3.44 \leq \alpha \leq 9.76$ (BCC), $9.76 \leq \alpha \leq 11.35$ (DQC), and $\alpha \geq 11.35$ (HEX).

This predicted phase behaviour for finite $c$ is very different from that of infinite $c$ as given by Lifshitz and Petrich\,\cite{lifshitz1997theoretical}. For infinite $c$, it was predicted that only the DDQC could become an equilibrium quasicrystalline phase. This behaviour is justified by the fact that when $c\rightarrow+\infty$, the Fourier wavevectors of DDQC can form the most triangles among the DDQC, DQC and OQC structures, thus decreasing the free energy of the DDQC structure. However, when $c$ is finite, more nonzero Fourier modes arise away from the circles with radii $1$ and $q$. The higher-harmonic contributions to the free energy could lower the free energy of the other quasicrystalline structures, resulting in the predicted phase behaviour from the current calculations. 

\begin{figure}[t]
	\centering
		\includegraphics[scale=0.32]{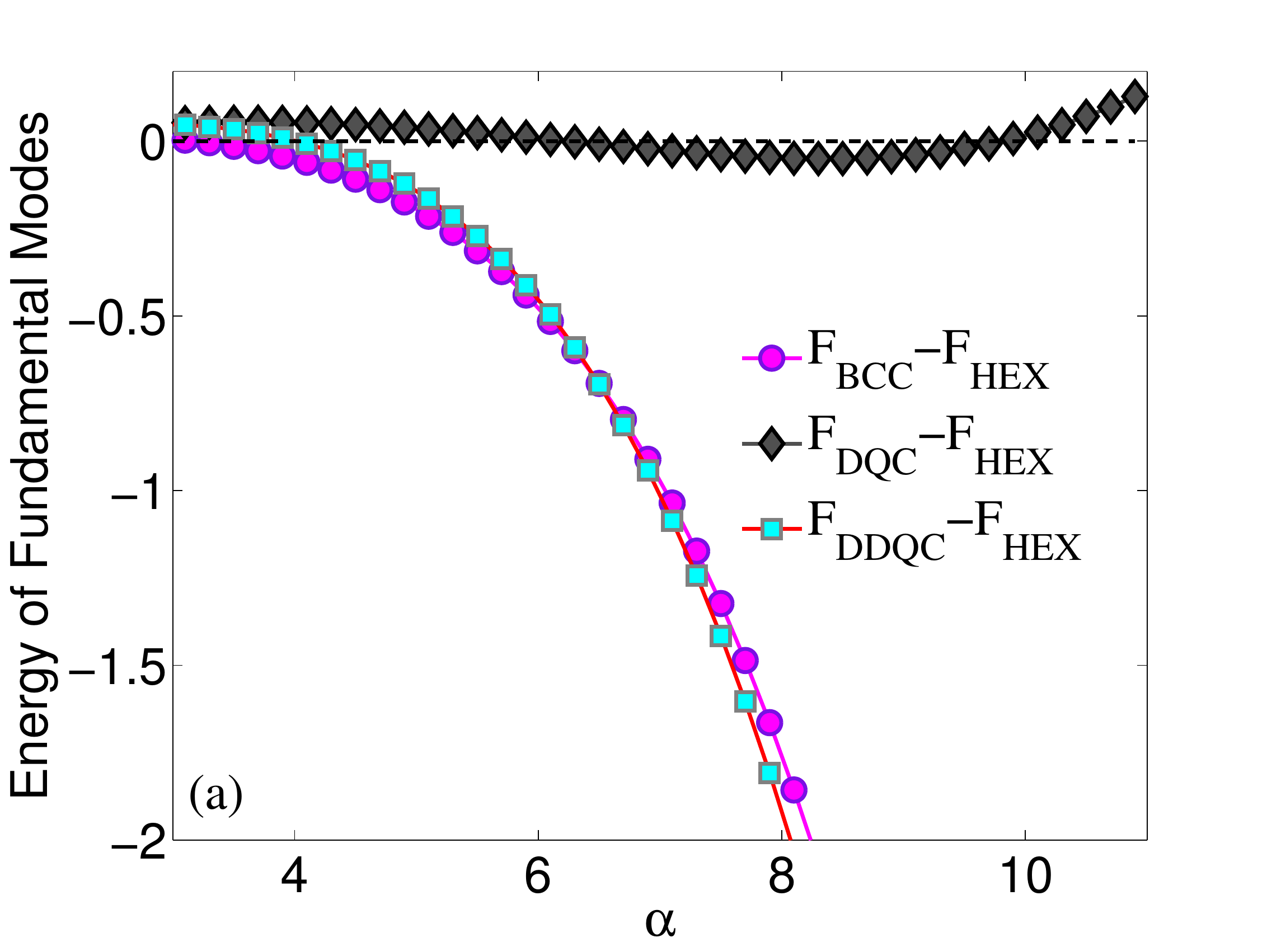}
		\includegraphics[scale=0.32]{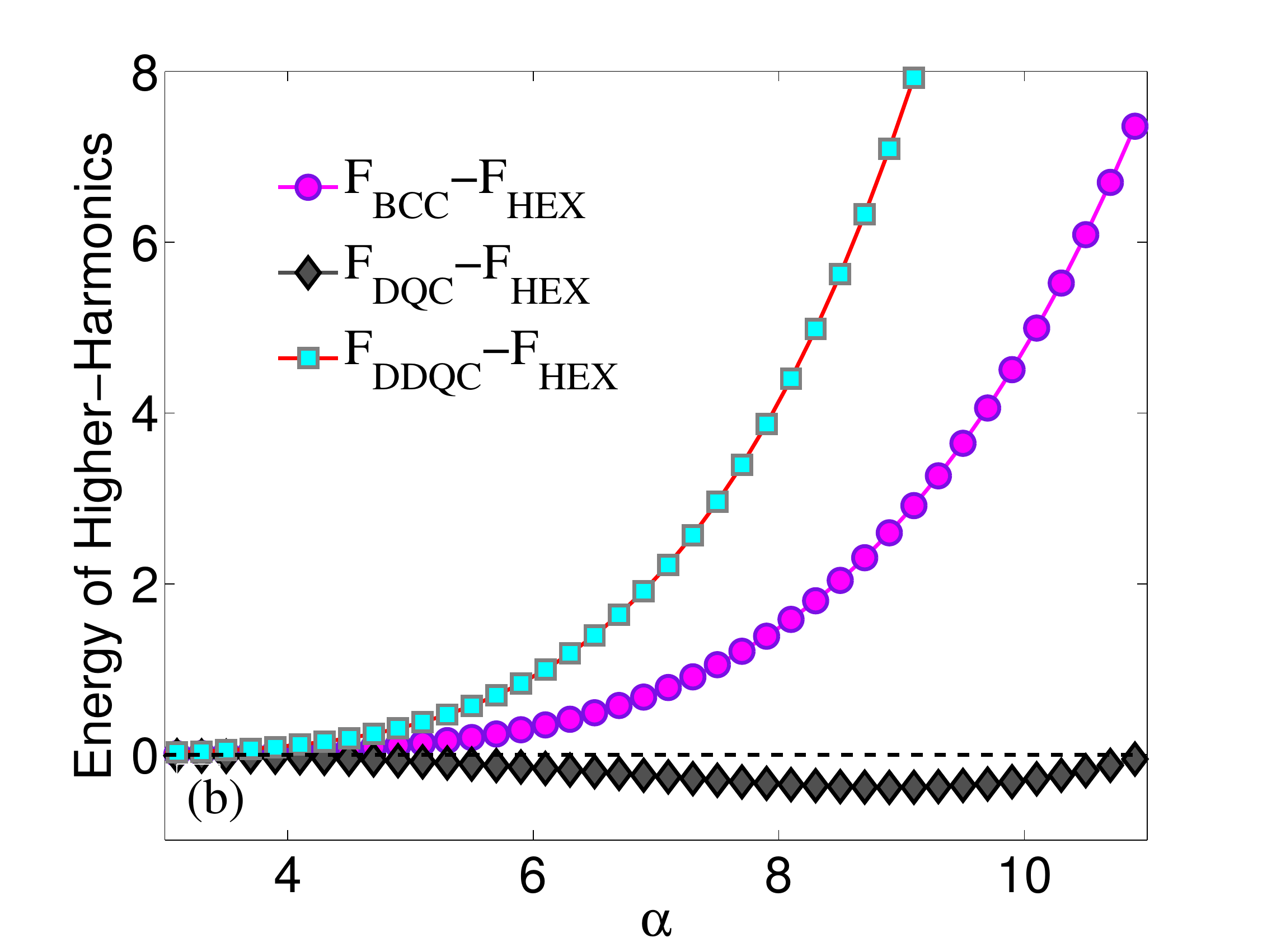}
	  \caption{The difference of (a) fundamental modes energy and
	  (b) higher-harmonic energy of various patterns from the
	  corresponding part of HEX as a function of $\alpha$ for
	  fixed $\varepsilon=-0.1$ and $c=100$.  The OQC is a metastable along
	  this phase path, therefore, its energy difference is not
	  shown. }
	\label{fig:energyplotxi-0.1Part}
\end{figure}

In order to analyze contributions to the free energy from the different Fourier modes, it is informative to divide the Fourier modes into two parts, {\it i.e.} fundamental modes and higher-harmonics. The fundamental modes are those Fourier modes with nonzero coefficients lying on circles with radius $1$ and $q$, and the higher-harmonics are the other Fourier modes with nonzero coefficients. The fundamental part of energy is defined by the contribution of the fundamental Fourier modes to the free energy, while the higher-harmonic part of energy is the remainder when subtracting the fundamental part of energy from the total energy.  The fundamental part of energy can be calculated analytically using the two-mode approximation method\,\cite{chaikin1995principles,lifshitz1997theoretical}. The analytic expressions of the fundamental part of energy corresponding to various patterns is given in the Appendix.

Fig.\,\ref{fig:energyplotxi-0.1Part} shows the fundamental and higher-harmonic contributions to the free energy for the DDQC, DQC and BCC phases using the HEX phase as the baseline. The free energy of the OQC structure is not included here since the OQC is always a metastable phase of the model system (see Fig.\,\ref{fig:energyplotxi-0.1}). From Fig.\,\ref{fig:energyplotxi-0.1Part} (a), it can be concluded that the DDQC is favoured by the fundamental modes because it has the largest number, $24$, of nonzero Fourier modes located on the circles with radii of $1$ and $q$, thus forming the most triangles in the Fourier space. On the other hand, since the energy penalty factor $c$ is finite, the higher-harmonics cannot be ignored. As Fig.\,\ref{fig:energyplotxi-0.1Part} (b) shows, these higher-harmonics have large impacts on the free energy of the system. HEX has the lowest higher-harmonics contribution to the free energy, since its Fourier modes with nonzero coefficients form the least tetragonal interaction. It can be observed from Eq.~\,(\ref{eqn:LPqcFour}) that the four-body interaction or the quartic term increases the free energy. Meanwhile, the differential term is no longer zero, which increases the energy. Due to the competition of the fundamental modes and higher-harmonics, the stable phases are DDQC, BCC, DQC and HEX patterns as $\alpha$ increases.

\begin{figure}[t]
	\centering
		\includegraphics[scale=0.25]{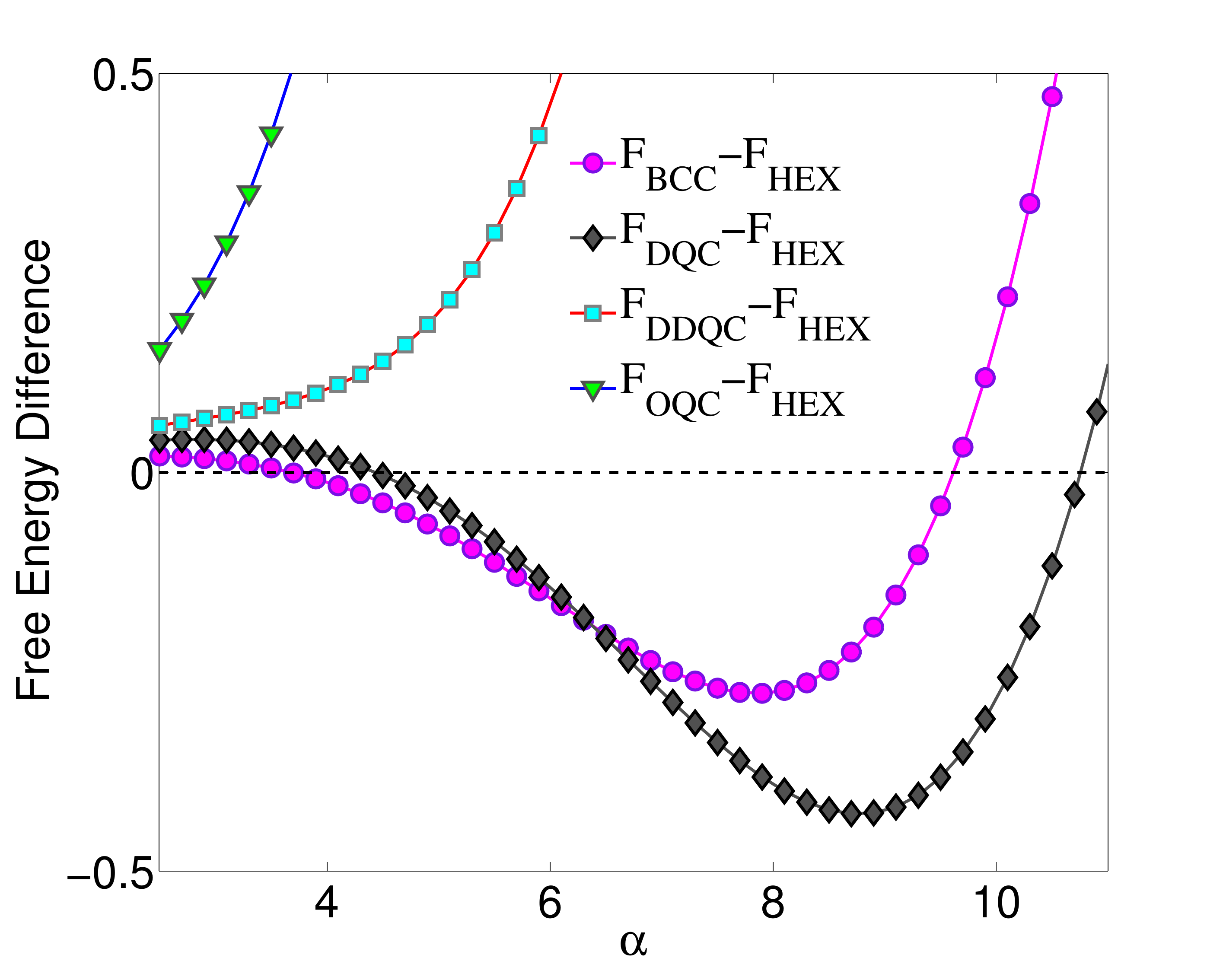}
	  \caption{The free energy differences from the value of
	  HEX as a function of $\alpha$ for fixed $\varepsilon=0.5$
	  and $c=100$.
	  }
	\label{fig:energyplotxi0.5}
\end{figure}

To further understand the influence of the higher-harmonics on the stability region of various phases, we set $\varepsilon=0.5$ and examine the contributions to the free energy as a function of $\alpha$. In this case, the quadratic and cubic terms in Eq.~\ref{eqn:LPqcFour} decrease the free energy since $\varepsilon>0$, while the quartic terms still increase the free energy. The free energy differences as a function of $\alpha\in[2.0, 11.0]$ are plotted in Fig.\,\ref{fig:energyplotxi0.5}.
For this set of parameters, the HEX, BCC and DQC structures form the possible equilibrium phases, whereas the DDQC and OQC structures become metastable. The stability regions of the BCC and DQC phases are $3.67 \leq\alpha \leq 6.38$ and $6.38 \leq \alpha \leq 10.75$, respectively. Compared to the previous discussion for the case of $\varepsilon = -0.1$, the stability region of the BCC structure becomes smaller. An interesting phenomenon is the reentrance transition of the HEX phases, in which its stability regions, $\alpha \leq 3.67$ and $\alpha \geq 10.75$, are separated by the BCC and DQC phases. Moreover, the HEX phase in these two regions share the same lattice spacing.

\begin{figure}[t]
	\centering
		\includegraphics[scale=0.32]{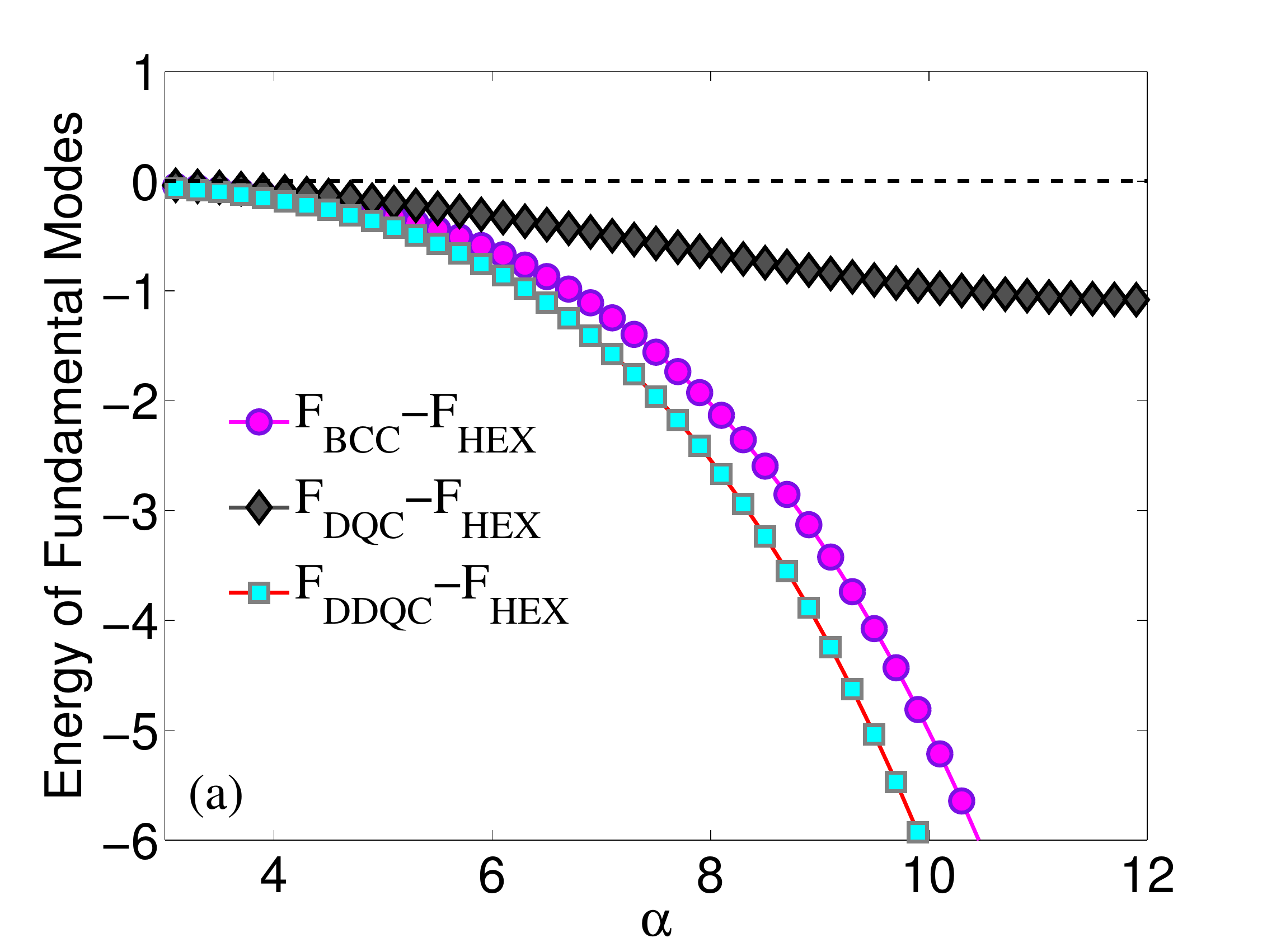}
		\includegraphics[scale=0.32]{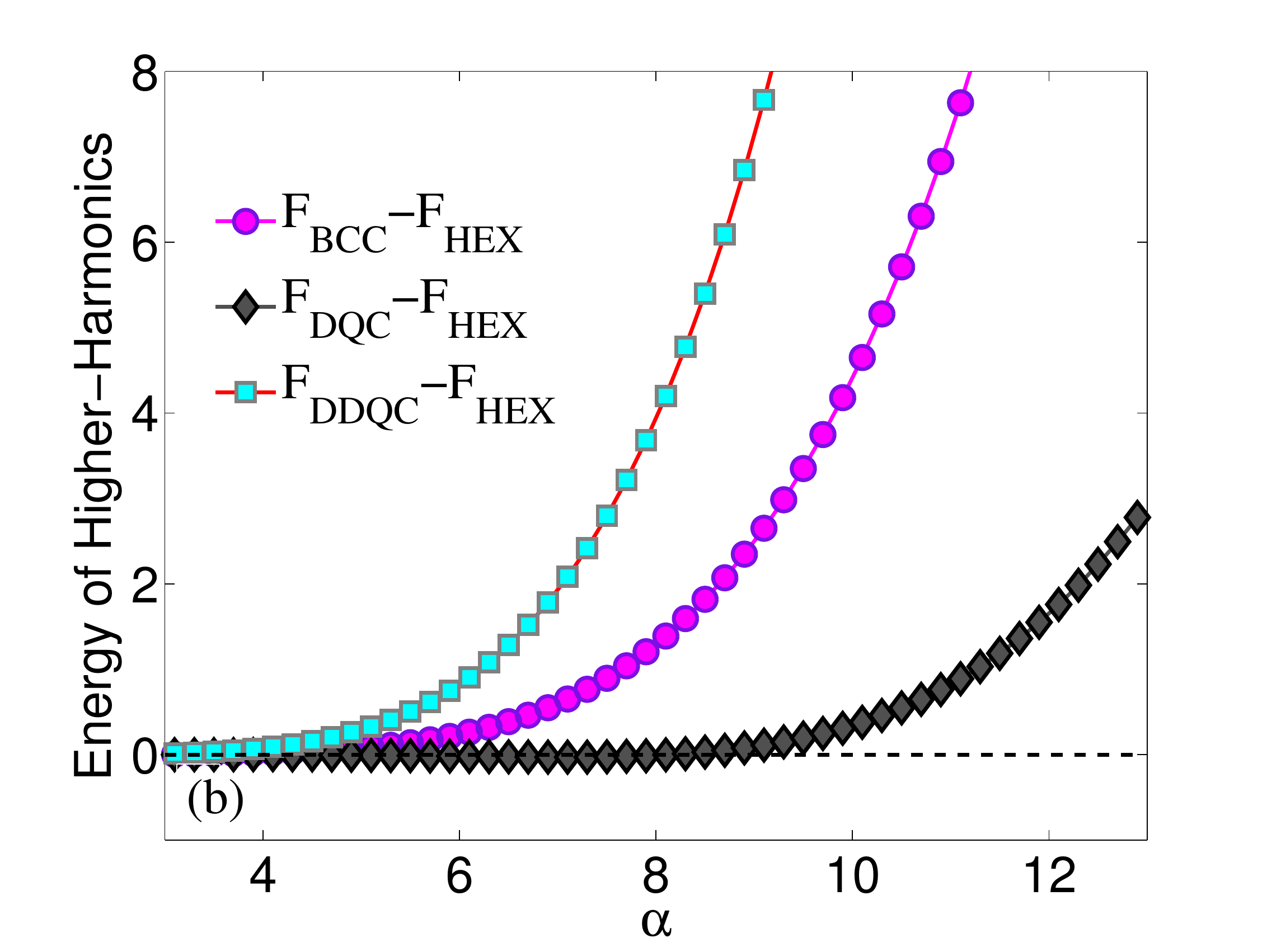}
	  \caption{The difference of (a) fundamental modes energy and (b)
	  higher-harmonic energy of different patterns from the
	  corresponding part of HEX as a function of $\alpha$ for
	  fixed $\varepsilon=0.5$ and $c=100$.  The OQC is a metastable along
	  this phase path, therefore, its energy difference is not
	  shown.  }
	\label{fig:energyplotxi0.5Part}
\end{figure}

Similar to the case of $\varepsilon = -0.1$, the contributions to the free energy from different Fourier modes can be used to understand the phase behavior (see Fig.\,\ref{fig:energyplotxi0.5Part}).
As is noted before, since the fundamental Fourier modes of DDQC and BCC structures can form more triangles than that of the HEX phase, the DDQC and BCC phases have lower free energy from the fundamental modes, as shown Fig.\,\ref{fig:energyplotxi0.5Part} (a). However, the free energy differences in both fundamental and higher-harmonic parts between the DQC and HEX phases exhibit non-monotonic behaviour. Specifically, when $\alpha$ is small, HEX has lower energy value in the fundamental part. As $\alpha$ increases, DQC becomes more favoured. When $\alpha$ is large enough, HEX has lower free energy again. From Eqs.~\eqref{eqn:swa:10} and \eqref{eqn:swa:6}, we find that DQC has more two-, three- and four-body interactions in the fundamental part of energy than the HEX does. More precisely, when $\alpha$ is small, the sum of the cubic terms in the energy expression of DQC is smaller than that of the HEX, and the sums of the quadratic and quartic terms are larger than the counterparts for HEX.  Therefore, when $\alpha$ is small, the quadratic term dominates the free energy; HEX thus has lower free energy than the DQC. As $\alpha$ increases, the cubic term becomes important in $F_{\mathrm{LP}}$, and thus DQC has lower energy in the fundamental part. When $\alpha$ is sufficiently large, the value of cubic and quartic terms of DQC becomes comparable with the HEX (results not shown). Thus the difference in the quadratic terms dominates the difference of the free energy in this case. Therefore, HEX becomes favored in energy again. The energy differences in the higher-harmonic part are given in Fig.\,\ref{fig:energyplotxi0.5Part} (b). The DDQC and BCC phases have relatively high energy as before. The energy difference between DQC and HEX has the similar behavior with the
fundamental part, which can be explained in a similar way.  Therefore, as $\alpha$ increases, the phase transition occurs in the sequence of HEX, BCC, DQC, and HEX.

\begin{figure}[t]
	\centering
		\includegraphics[scale=0.25]{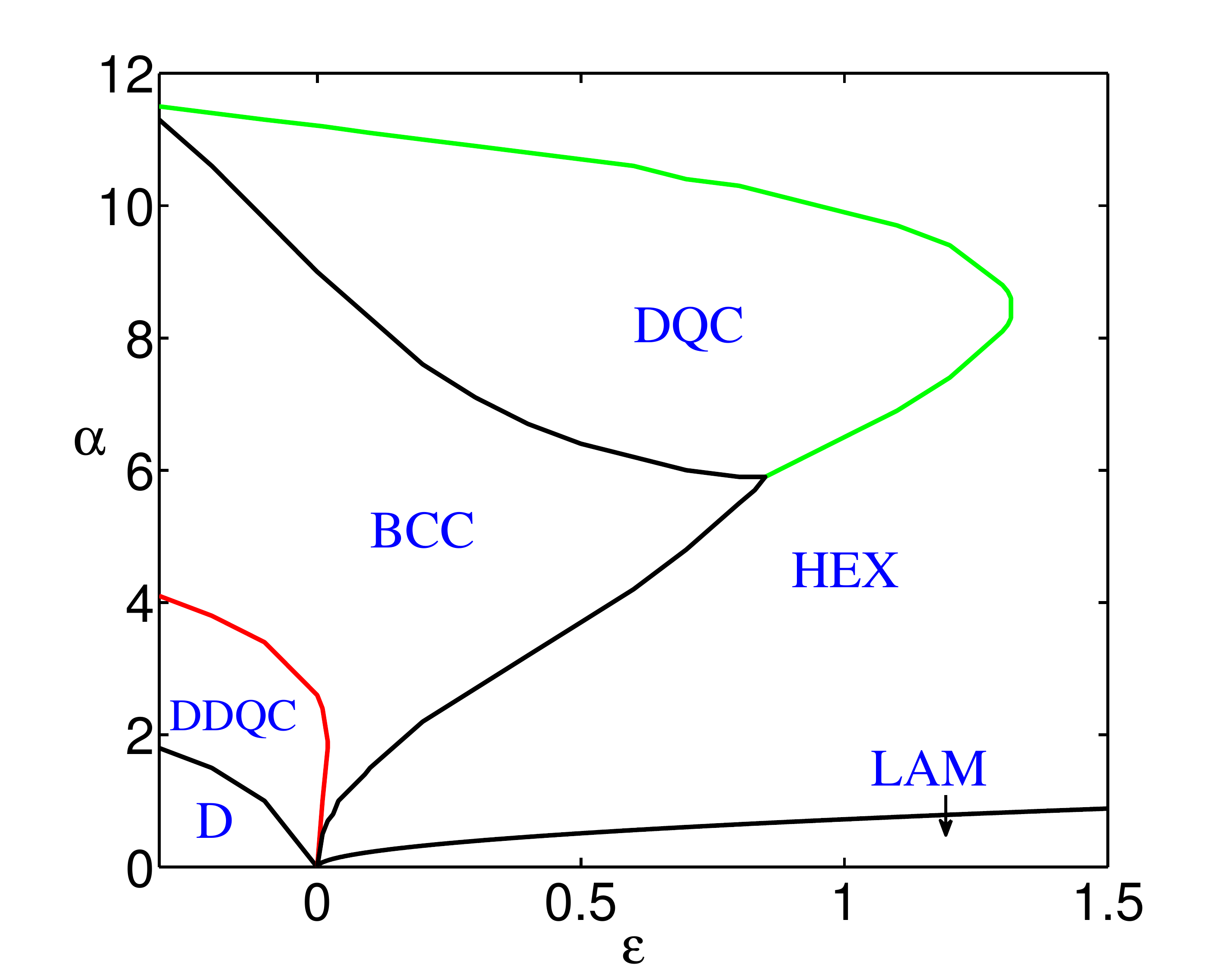}
	  \caption{Phase diagram of the LP model with
	  $c=100$. LAM and D represent the 2-fold pattern and
	  disordered phase, respectively.
	  For DDQC and DQC, the structural parameter $q$ equals
	  to $2\cos(\pi/12)$ and $2\cos(\pi/5)$, respectively. For
	  periodic crystals, $q=2\cos(\pi/5)$.}
	\label{fig:phaseDiagram}
\end{figure}

The phase transition sequence for other values of $\varepsilon$ can be obtained by repeating the free energy comparison among the candidate structures. The results of the phase transition sequences can be summarized in terms of phase diagrams in the $\varepsilon$-$\alpha$ plane. For a fixed value of $c=100$, the phase diagram in the range of $-0.3\leq\varepsilon \leq 1.5$  and $0\leq \alpha \leq 12$ is presented in Fig.\,\ref{fig:phaseDiagram} for the LP free energy functional.
Besides the DDQC, DQC, OQC, HEX and BCC phases discussed above, two more phases, the 2-fold symmetric pattern, or the lamellar phase (LAM), and the disordered phase (D) are included in the phase diagram. The regions of stability of the different phases are obtained by comparing the free energy of these seven candidate structures. The phase boundaries are determined by calculating the cross over point of the free energies of the two neighbouring phases. Because the contribution from the higher-harmonics becomes significant for the cases with finite values of $c$, the predicted phase diagram from the current study includes more features when compared with the previous phase diagram obtained from the analytic calculation using the two-mode approximation\,\cite{lifshitz1997theoretical}.

In the previous work, Lifshitz and Petrich used the two-mode approximation method\,\cite{chaikin1995principles} to analytically analyze the phase behavior of the LP model when $c$ is sufficiently large. It is found that when $q=2\cos(\pi/12)$ and $c\rightarrow+\infty$, the LAM phase is stable for $\varepsilon/\alpha^2>1.91313$; the HEX phase has the lowest energy for $0.08776<\varepsilon/\alpha^2<1.91313$; and the DDQC is stable for $\varepsilon/\alpha^2<0.08776$. While the generic feature of the phase diagrams is maintained, there are significant differences between the analytic phase diagram and the numerical result presented in the current work, which contains more features by taking higher-harmonics into account.  The first noticeable difference is that the current phase diagram includes the DQC as a new equilibrium quasicrystal phase, whereas the previous phase diagram only contains the DDQC as the unique stable quasicrystalline phase. It is interesting to notice that the DQC-HEX phase boundary possess a special turning point at $(\varepsilon, \alpha)=(1.32,8.46)$. Increasing $\alpha$ at $\varepsilon < 1.32$ leads to a reentrance transition of the HEX phase. Another interesting feature is the prediction of a triple point at $(\varepsilon, \alpha) =(0.85, 5.9)$, at which the BCC, DQC and HEX phases coexist. The OQC is only observed as a metastable phase, which is in agreement with the analytic result by two-mode approximation. Moreover, a three-dimensional phase with spheres packed on a body-centred-cubic lattice (BCC) is included in the current phase diagram.

\section{Summary}
\label{sec:sum}

In summary, we have applied the projection method to obtain quasicrystalline structures as possible local minima of the LP free energy functional. The projection method enables accurate numerical calculations of the free energy of the quasicrystals, resulting in a phase diagram with more accurate phase boundaries and, most importantly, more ordered phases as compared with the previous study. Specifically, three two-dimensional quasiperiodic patterns, namely the OQC, DQC and DDQC structures, and several periodic phases (LAM, HEX and BCC) emerge from our calculations as candidate structures for the construction of the phase diagram. By comparing the free energy of the candidate structures, a phase diagram for the case of $c = 100$ is obtained in the $\varepsilon$-$\alpha$ place. It is predicted that for the LP free energy function given in Eq.\ref{eqn:LP}, the 10-fold (DQC) and 12-fold (DDQC) symmetric quasicrystals can become equilibrium phases in the phase diagram.  On the other hand the 8-fold (OQC) symmetric quasicrystal is found to be a metastable phase of the model system. The prediction of a stable DQC phase within the LP free energy functional is a new result. The present study complements and extends the previous work of Lifshitz and Petrich\,\cite{lifshitz1997theoretical}. 

The two-mode approximation provides an accurate description of the LP model at the limit of $c = +\infty$, at which the wavevectors are forced to be on two circles defined by $|\mathbf{k}| = 1$ and $|\mathbf{k}| = q$. A finite value of $c$, as exemplified by the case of $c=100$, allows the occurrence of higher-harmonics other than the modes with $|\mathbf{k}| = 1$ and $|\mathbf{k}| = q$.  The additional Fourier modes lead to phase diagram at finite $c$ that is significantly different from the phase behavior of the case of $c = +\infty$.  In particular, the quasicrystalline phases DQC, DDQC, the lamellar phase, HEX and BCC phase are predicted to be potentially stable phases for the LP free energy functional.  These results provide a good understanding of the rich phase behaviour contained in the simple LP model. The numerical methods and insights obtained from them can be helpful for further studying more complex soft matter systems such as block copolymers and soft particles with multi-length-scale interactions.

\section{Acknowledgements}
The work is supported by the National Science Foundation of China (Grant No.\,11421110001, 11421101, 21274005, and 11401504).  KJ would like to thank the financial supported by the Hunan Science Foundation of China (Grant No.~2015JJ3127).  ACS acknowledges the support from the Natural Sciences and Engineering Research Council (NSERC) of Canada.

\
\appendix
\section{Free Energy of fundamental Fourier Modes}
\label{subsec:swa}

In the two-mode approximation valid in the limit of $c = +\infty$, the Fourier modes are restricted to the two circles defined by $|\mathbf{k}| = 1$ and $|\mathbf{k}| = q$. Keeping only the Fourier coefficients at $|\mathbf{k}| = 1$ and $|\mathbf{k}| = q$, the free energy of the system can be obtained analytically. The analytic expressions of fundamental energy (which has been defined in Section \ref{subsec:Relative Stability of Ordered Phases and Phase diagram}) of different patterns are given as follows.
\begin{equation}
\begin{aligned}
	F_{\mathrm{DDQC,L}} = &
	- 6\varepsilon^ (\hat{\phi}_1^2+\hat{\phi}_q^2)
	-24\alpha(\hat{\phi}_1^2\hat{\phi}_q+\hat{\phi}_1 \hat{\phi}_q^2)
	\\
	& -8\alpha(\hat{\phi}_1^3+\hat{\phi}_q^3)
	+99(\hat{\phi}_1^4+\hat{\phi}_q^4)
	\\
	& +144(\hat{\phi}_1 \hat{\phi}_q^3+\hat{\phi}_1^3\hat{\phi}_q) +
	360\hat{\phi}_1^2\hat{\phi}_q^2,
	\label{eqn:swa:12}
\end{aligned}
\end{equation}
\begin{equation}
\begin{aligned}
	F_{\mathrm{DQC,L}} = &
	-5\varepsilon(\hat{\phi}_1^2 + \hat{\phi}_q^2)
	 -20\alpha(\hat{\phi}_1^2\hat{\phi}_q+\hat{\phi}_1\hat{\phi}_q^2)
	 \\ &
	 +\frac{15}{2}(9\hat{\phi}_1^4+8\hat{\phi}_1^3\hat{\phi}_q+28\hat{\phi}_1^2\hat{\phi}_q^2
	+ 8\hat{\phi}_1\hat{\phi}_q^3 +9\hat{\phi}_q^4),
	\label{eqn:swa:10}
\end{aligned}
\end{equation}
\begin{equation}
\begin{aligned}
	F_{\mathrm{OQC,L}} = &
	-4\varepsilon(\hat{\phi}_1^2 + \hat{\phi}_q^2)
	-16\alpha\hat{\phi}_1^2\hat{\phi}_q
	\\
	& +6(7\hat{\phi}_1^4+24\hat{\phi}_1^2\hat{\phi}_q^2+7\hat{\phi}_q^4),
	\label{eqn:swa:8}
\end{aligned}
\end{equation}
\begin{align}
	F_{\mathrm{HEX,L}} =
	-3\varepsilon \hat{\phi}_{1}^2
	-4\alpha\hat{\phi}_{1}^3 +
	\frac{45}{2}\hat{\phi}_{1}^4,
	\label{eqn:swa:6}
\end{align}
\begin{align}
	F_{\mathrm{BCC,L}} =
	-12\varepsilon \hat{\phi}_{1}^2
	-16\alpha\hat{\phi}_{1}^3 +
	135\hat{\phi}_{1}^4
	\label{eqn:swa:BCC},
\end{align}
where $\hat{\phi}_1, \hat{\phi}_q\in\mathbb{R}$ stand for the Fourier coefficients on the circles of $1$ and $q$, respectively.

\end{document}